\documentclass[
 reprint,
 amsmath,amssymb,
 aps,
 floatfix,
]{revtex4-2}

\usepackage[colorlinks=true,citecolor=blue,urlcolor=blue,linkcolor=blue]{hyperref}

\usepackage{amsmath,amssymb,bm}
\usepackage{graphicx}
\usepackage{xcolor}
\usepackage{hyperref}

\hypersetup{colorlinks=true,citecolor=blue,linkcolor=blue,urlcolor=blue}

\newcommand{\gone}{g^{(1)}}

\begin{document}

\title{Three-Dimensional Kardar--Parisi--Zhang Scaling in Polariton Condensates}

\author{Junhui Cao$^{1}$}%
\author{Denis Novokreschenov$^{1}$}%
\author{Artem Alexandrov$^{2,3}$}%
\author{Timothy Halpin-Healy$^{4}$}%
\email{healy@phys.columbia.edu}
\author{Alexey Kavokin$^{1,5,6,7}$}%
\email{a.kavokin@westlake.edu.cn}
\affiliation{$^{1}$Abrikosov Center for Theoretical Physics, Moscow Center for Advanced Studies, Moscow 141701, Russia\\
$^{2}$Phystech School of Applied Mathematics and Computer Science, Moscow Institute of Physics and Technology, Dolgoprudny 141700, Russia\\
$^{3}$Steklov Mathematical Institute of Russian Academy of Sciences, Moscow 119333, Russia\\
$^{4}$Physics Department, Barnard College, Columbia University, New York, New York 10027, USA\\
$^{5}$School of Science, Westlake University, Hangzhou 310024, Zhejiang Province, China\\
$^{6}$Department of Physics, St. Petersburg State University, University Embankment, 7/9, St. Petersburg, 199034, Russia\\
$^{7}$Russian Quantum Center, Skolkovo, Moscow Region, 121205, Russia
}

\date{\today}

\begin{abstract}
Kardar--Parisi--Zhang (KPZ) universality provides an example of macroscopic scaling generated by microscopic violation of detailed balance. While one- and two-dimensional realizations have been explored in driven condensates and growing interfaces, demonstrating KPZ scaling in three spatial dimensions remains a major challenge. Here we propose a three-dimensional exciton-polariton crystal as a platform for observation of 3D KPZ universality. Starting from a stochastic driven-dissipative Gross-Pitaevskii equation for a condensate formed in a three-dimensional photonic-crystal lower-polariton band, we eliminate the massive density and reservoir modes and obtain an effective $3+1$-dimensional KPZ equation for the condensate phase. Numerical simulations of both the KPZ equation and the full driven-dissipative polariton model show an intermediate-asymptotic regime in which the first-order coherence obeys $-\ln |\gone(0,\Delta t)|\propto |\Delta t|^{2\beta}$ and $-\ln |\gone(\Delta r,0)|\propto |\Delta r|^{2\chi}$, with exponents consistent with the $3+1$ KPZ benchmarks $\beta= 0.1845$, $\chi= 0.3135$. Our results identify three-dimensional polariton crystals as a controllable quantum fluid route to higher-dimensional nonequilibrium universality.
\end{abstract}

\maketitle

\paragraph*{Introduction.---}
In systems close to thermal equilibrium, universality usually means that long-wavelength fluctuations are organized by symmetries, conservation laws, and dimensionality \cite{hohenberg1977theory,odor2004universality,mukherjee2021conserved}. Driven-dissipative condensates are different. They continuously exchange particles and energy with their environment, and their steady states are not obtained by minimizing an equilibrium free energy. An important question is therefore whether their macroscopic coherence is governed by equilibrium universality or by nonequilibrium fixed points \cite{PhysRevLett.101.041603, chantesana2019kinetic}.

The Kardar--Parisi--Zhang (KPZ) equation \cite{PhysRevLett.56.889, SNS,takeuchi2018appetizer} has taken center-stage as the canonical nonlinear stochastic partial differential equation for such far-from-equilibrium scaling,
\begin{equation}
  \partial_t h({\bf r},t)=\nu \nabla^2 h+\frac{\lambda}{2}(\nabla h)^2+\sqrt D \eta({\bf r},t),
  \label{eq:kpz_intro}
\end{equation}
where $h$ is the height field, $\eta$ is short-range spatiotemporally uncorrelated noise, $\langle \eta(\mathbf r,t)\eta(\mathbf r',t')\rangle=2\delta^{(3)}(\mathbf r-\mathbf r')\delta(t-t')$, while $\nu$, $\lambda$ and $D$ are phenomenological parameters. In a condensate, the same equation can emerge for the phase $\theta$ of the order parameter. Here, the phase plays the role of the KPZ height, and the nonlinear term $(\nabla\theta)^2$ is allowed because drive and dissipation break the equilibrium constraints that would otherwise forbid it or mask the asymptotic KPZ regime in realistic experiments at long wavelengths. This mapping has made exciton-polariton condensates one of the most promising platforms for observing KPZ physics in a low-dimensional, phase-coherent quantum fluid \cite{sieberer2013dynamical,fontaine2022kardar,widmann2026observation,ym2g-t8vj,doi:10.1126/science.aeg3150}.

The step from one or two to three spatial dimensions is not a straightforward extension. In $d=1$, KPZ scaling is exceptionally robust and exactly constrained. In $d=2$, the KPZ nonlinearity is marginally relevant, yet driven condensates can display a clear departure from the equilibrium Berezinskii--Kosterlitz--Thouless picture \cite{berezinskii1971destruction,kosterlitz1973ordering,kosterlitz1974critical}. In $d=3$, however, the situation is more demanding. First, the KPZ equation has a weak-coupling smooth phase and a strong-coupling rough phase \cite{TangI}. A three-dimensional experiment must be driven into the latter one, instead of showing diffusive Edwards--Wilkinson phase relaxation \cite{edwards1982surface}. Second, the growth exponent is small 
\cite{TangII}, $\beta\approx 0.18$, so the observable coherence decay changes slowly with time and requires a broad scaling window. Third, finite-size saturation and microscopic transients are more severe because a three-dimensional sample contains more transverse modes and has multiple relaxation channels. Direct numerical studies of the $3+1$ KPZ equation have themselves required large-scale simulations, with estimates converging only gradually from early values
\cite{ala1993scaling,moser1994vectorized} such as $\beta=0.180(5),0.181(7)$ to recent high-precision benchmarks \cite{halpin2025kardar} which land on $\beta=0.1845(4)$. A physical realization must solve the corresponding problem. It must provide a large, phase-coherent, robust three-dimensional driven condensate with tunable noise, nonlinearity, loss, and effective mass.

Here we argue that a three-dimensional polariton crystal provides precisely this setting \cite{cao2026bose}. The platform we have in mind is an inverse-opal photonic crystal doped with excitonic emitters, such as semiconductor quantum dots. Strong light-matter coupling converts a selected photonic Bloch band into lower and upper polariton bands. Unlike a nanowire or planar microcavity, where the condensate is essentially one- or two-dimensional \cite{jia2026femtosecond,altman2015two}, a photonic crystal supports a three-dimensional Bloch dispersion with well-defined valleys, curvatures, and density-of-states structure \cite{john1987strong,yablonovitch1987inhibited}. In the relevant design, the lower polariton has a tunable global minimum at the high-symmetry point of the face-centered-cubic Brillouin zone. This band-structure control allows one to choose a valley with approximately isotropic curvature, engineer the phase stiffness, and tune the driven-dissipative KPZ coupling without changing the underlying dimensionality.

In this paper, we start from a stochastic open-dissipative Gross--Pitaevskii equation for a polariton condensate in a three-dimensional Bloch band. By eliminating the massive condensate-density and reservoir-density fluctuations, we derive an effective anisotropic KPZ equation for the condensate phase. We then show numerically that the KPZ equation gives the expected $3+1$ benchmark scaling, and that the full polariton model displays the same exponents in an intermediate-asymptotic regime. Finally, we discuss how the predicted scaling can be measured through the first-order correlation function.


\paragraph*{Driven-dissipative polariton model.---}
We consider a condensate formed near a lower-polariton valley of a three-dimensional photonic crystal. On length scales large compared with the lattice constant, the lower-polariton field is described by a valley-envelope order parameter $\psi({\bf r},t)$ with anisotropic effective masses $m_i$ $(i=x,y,z)$. For nonresonant pumping, the condensate is coupled to an incoherent reservoir density $n_R({\bf r},t)$. In units with $\hbar=1$, the stochastic driven-dissipative Gross--Pitaevskii equation reads
\begin{equation}
\begin{aligned}
i\partial_t\psi =\Bigg[&-\sum_{i=x,y,z}\frac{1}{2m_i}\partial_i^2 +g|\psi|^2+g_R n_R+\\
&\frac{i}{2}(R n_R-\gamma_c)\Bigg]\psi+\xi_\psi,
\end{aligned}
\label{eq:gpe} 
\end{equation}
\begin{equation}
    \partial_t n_R ={} P-(\gamma_R+R|\psi|^2)n_R+D_R\nabla^2 n_R+\xi_R.
    \label{eq:reservoir} 
\end{equation}
Here $g$ is the polariton-polariton interaction, $g_R$ is the reservoir-induced blueshift, $R$ is the stimulated scattering rate from the reservoir to the condensate, $\gamma_c$ and $\gamma_R$ are the condensate and reservoir decay rates, and $P$ is the pump. The noise term represents fluctuations associated with pumping, decay, and reservoir relaxation. The microscopic values of the coefficients are nonuniversal and depend on the excitonic fraction, radiative leakage, and the photonic-crystal band curvature. Their role in the scaling theory is to determine whether the effective phase dynamics lies in the weak-coupling or strong-coupling KPZ regime.

For a spatially uniform pump above threshold, the mean-field steady state is
\begin{equation}
 n_R^0=\frac{\gamma_c}{R},\qquad
 \rho_0=\frac{P-P_{\rm th}}{\gamma_c},\qquad
 P_{\rm th}=\frac{\gamma_c\gamma_R}{R},
 \label{eq:steady}
\end{equation}
with oscillation frequency
\begin{equation}
 \omega_0=g\rho_0+g_R n_R^0.
 \label{eq:omega0}
\end{equation}
We write
\begin{align}
 \psi({\bf r},t)&=\sqrt{\rho_0+\epsilon({\bf r},t)}
 \exp[-i\omega_0 t+i\theta({\bf r},t)],
 \\
 n_R&=n_R^0+\delta n_R,
 \label{eq:phase_density}
\end{align}
where $\epsilon$ and $\delta n_R$ are density fluctuations and $\theta$ is the condensate phase. The density and reservoir modes are massive, while the phase is the only Goldstone mode. At long wavelengths and low frequencies, $\epsilon$ and $\delta n_R$ can be eliminated. Keeping the leading gradient terms gives an effective phase equation
\begin{equation}
 \partial_t\theta=
 \sum_i \nu_i\partial_i^2\theta+\frac{1}{2}\sum_i\lambda_i(\partial_i\theta)^2+\eta_{\rm eff}+O(\nabla^4),
 \label{eq:anisotropic_kpz}
\end{equation}
where $\eta_{\rm eff}$ is a short-range effective noise. In a local adiabatic approximation, defining $\Gamma_R=\gamma_R+R\rho_0$, one finds
\begin{equation}
 \nu_i\approx \frac{1}{m_i}\left(\frac{g\Gamma_R}{R\gamma_c}-\frac{g_R}{R}\right),
 \qquad
 \lambda_i\approx -\frac{1}{m_i}.
 \label{eq:coefficients}
\end{equation}
When the effective valley is close to isotropic, Eq.~\eqref{eq:anisotropic_kpz} reduces to the standard $3+1$ KPZ equation for $\theta$,
\begin{equation}
\partial_t\theta=\nu\nabla^2\theta+\frac{\lambda}{2}(\nabla\theta)^2+\eta_{\rm eff}.
 \label{eq:3d_kpz}
\end{equation}
The finite dimensionless KPZ coupling is controlled experimentally by the pump distance from threshold, the effective mass, the noise strength, and the decay rates. This tunability is one of the main advantages of the polariton-crystal platform.

\paragraph*{First-order correlation.---}
The observable quantity is the first-order correlation function
\begin{equation}
 \gone(\Delta{\bf r},\Delta t)=
 \frac{\langle \psi^*({\bf r},t)\psi({\bf r}+\Delta{\bf r},t+\Delta t)\rangle}
 {\sqrt{\langle |\psi({\bf r},t)|^2\rangle\langle |\psi({\bf r}+\Delta{\bf r},t+\Delta t)|^2\rangle}}.
 \label{eq:g1_def}
\end{equation}
In the phase-dominated regime, for Gaussian phase fluctuations with zero mean,
\begin{equation}
\begin{aligned}
 -\ln |\gone(\Delta{\bf r},\Delta t)|
 &\approx\frac{1}{2}\left\langle (\Delta\theta-\left\langle\Delta\theta\right\rangle)^2\right\rangle\\
 &\approx\frac{1}{2}
 \left\langle [\theta({\bf r}+\Delta{\bf r},t+\Delta t)-\theta({\bf r},t)]^2\right\rangle.
\end{aligned}
 \label{eq:g1_phase}
\end{equation}
The KPZ fixed point predicts the scaling form for the stationary-state space-time correlator:
\begin{equation}
 -\ln |\gone(\Delta r,\Delta t)|
 \propto|\Delta t|^{2\beta}
 F\!\left(\frac{\Delta r}{|\Delta t|^{1/z}}\right),
 \label{eq:g1_scaling}
\end{equation}
with
\begin{equation}
 -\ln |\gone(0,\Delta t)|\propto |\Delta t|^{2\beta},
 \qquad
 -\ln |\gone(\Delta r,0)|\propto |\Delta r|^{2\chi}.
 \label{eq:g1_powerlaws}
\end{equation}

For the isotropic $3+1$ KPZ universality class, the best direct numerical estimates for the temporal
\cite{halpin2025kardar} and spatial
\cite{Enzo} exponents reveal, resp., 
\begin{equation}
 \beta= 0.1845(4),
 \qquad
 \chi= 0.3135(15)
 \label{eq:3d_exponents}
\end{equation}
which yields the dynamic exponent $z=\chi/\beta=1.699(12)$ consistent with the fundamental KPZ identity $\chi+z=2$, which follows from the scale-invariant, nonrenormalized nature of the central KPZ parameter $\lambda$.


\begin{figure}[t]
\centering
\includegraphics[width=\linewidth]{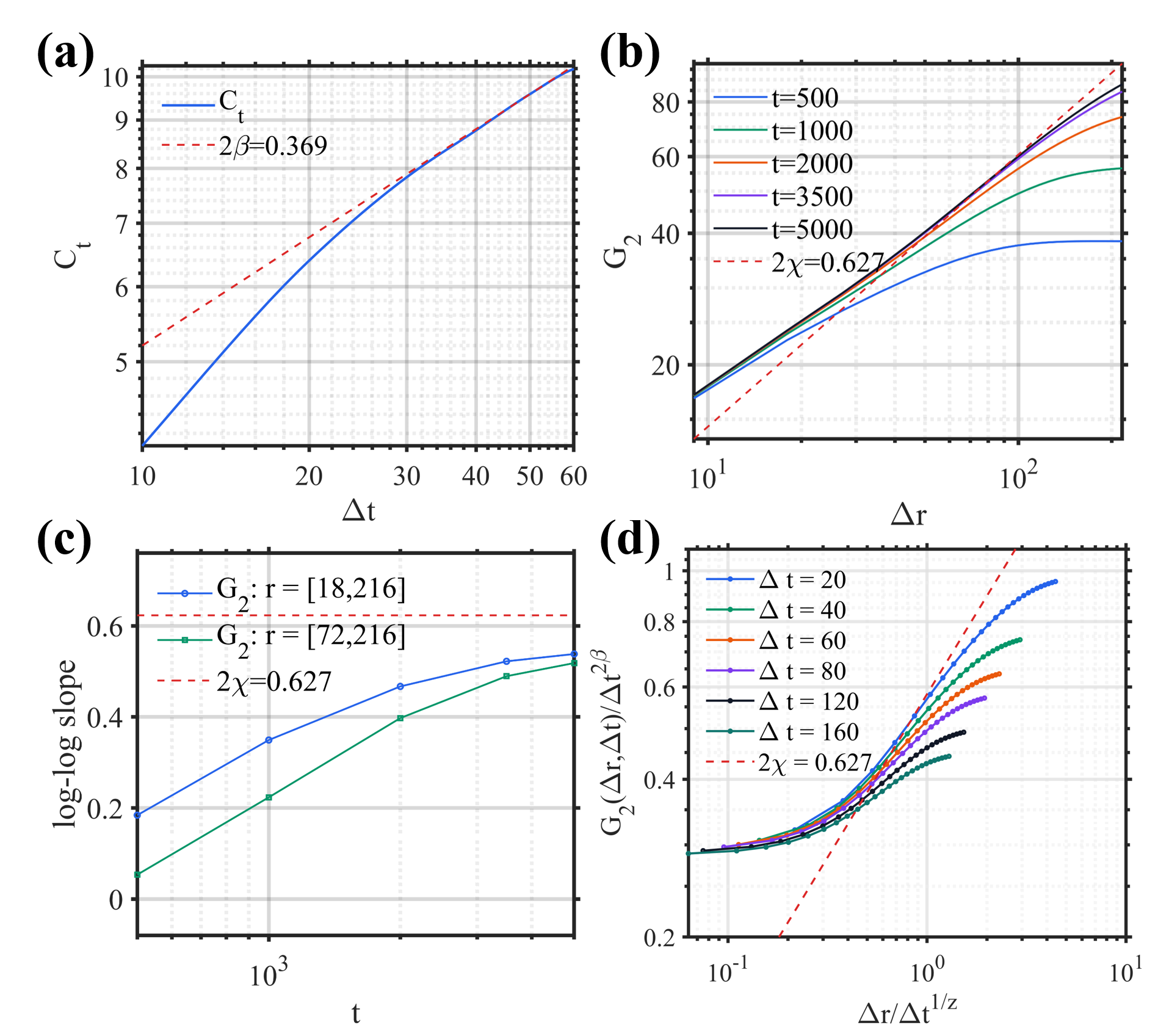}
\caption{$3+1$ KPZ benchmark in the strong-coupling regime. The main run uses $L=128$, $dx=9$, $\nu=0.9$, $dt=0.5$, $g_{\rm eff}=500$, and $10^4$ noise realizations. (a) Temporal compact coherence $C_t=-\log |g^{(1)}(0,\Delta t)|$ with the $2\beta=0.369$ guide, while the short-lag fit $40\le\Delta t\le60$ gives slope $0.37$. (b) Equal-time height correlator $G_2(\Delta r,t)=\langle[\theta({\bf r}+{\bf \Delta r},t)-\theta({\bf r},t)]^2\rangle$. (c) Spatial slopes versus observation time. (d) Dynamical scaling of the spatiotemporal correlator. The curves for $\Delta t=20,40,60,80,120$ and $160$ are plotted as $G_2(\Delta r,\Delta t)/\Delta t^{2\beta}$ versus $\Delta r/\Delta t^{1/z}$.}
\label{fig:analytic}
\end{figure}

\paragraph*{Numerical evidence for $3+1$ KPZ scaling.---}
We first simulate Eq.~\eqref{eq:3d_kpz} as a control calculation. This model isolates the universal strong-coupling fixed point without complications from density relaxation, reservoir transients, or finite polariton lifetime. Starting from a flat phase configuration, we integrate the stochastic equation on cubic lattices with periodic boundary conditions and evaluate the phase correlation function after averaging over noise realizations.
The numerically useful coupling of this KPZ problem is
\begin{equation}
 g_{\rm eff}=dx^3\frac{\lambda^2D}{\nu^3}.
 \label{eq:geff}
\end{equation}
For $dx=9$, $\lambda=1$, and $D=0.5$, the commonly used $\nu=1$ point gives $g_{\rm eff}=364.5$ \cite{moser1994vectorized}. Lowering $\nu$ increases the coupling but also makes the explicit Euler update more vulnerable to rare large gradients. We found that $\nu=0.9$ with $dt=1$ can be seed-dependent, while reducing the step to $dt=0.5$ stabilizes the same strong-coupling point, $g_{\rm eff}=500$, without changing the target continuum coupling. All runs discussed below use $L=128$, $dx=9$, $\nu=0.9$, $dt=0.5$, and evolve to $t=5000$, with $10^4$ noise realizations averaged.

Figure~\ref{fig:analytic} summarizes our numerical study of the 3+1 KPZ equation Eq.~\eqref{eq:3d_kpz} in the stationary-state, the regime  of relevance to the exciton-polariton experiments.
The temporal coherence, illustrated in Fig.~\ref{fig:analytic}a, captures the later time characteristic of the KPZ stationary-state statistics and gives a short-lag exponent
\begin{equation}
 -\ln |\gone(0,\Delta t)|\propto \Delta t^{0.37}
 \quad (40\le\Delta t\le60),
 \label{eq:fitted_phase_temporal}
\end{equation}
close to the benchmark value $2\beta=0.369$, before the compact correlator saturates at larger lags.

For the spatial correlations of the pure KPZ problem, we use the noncompact equal-time height correlator
\begin{equation}
 G_2(\Delta r,t)=
 \left\langle [\theta({\bf r}_0+{\bf \Delta r},t)-\theta({\bf r}_0,t)]^2\right\rangle.
 \label{eq:g2_noncompact}
\end{equation}
Unlike the compact equal-time coherence, which is already saturated in the relevant distance range, $G_2$ retains the growth of the unwrapped KPZ field. At $t=5000$, see Fig.~\ref{fig:analytic}b, the averaged fit gives
\begin{equation}
 G_2(\Delta r,t=5000)\propto \Delta r^{0.58}
 \quad (48\le \Delta r\le108).
 \label{eq:fitted_phase_spatial}
\end{equation}
As shown in Fig.~\ref{fig:analytic}c, the ensemble value is more conservative, but still shows clear movement toward the $2\chi=0.627$ benchmark as time increases.

A more complete test is provided by the noncompact spatiotemporal height correlator,
\begin{equation}
G_2(\Delta r,\Delta t)=\left\langle\left[\Delta\theta-\langle\Delta\theta\rangle\right]^2\right\rangle ,
\end{equation}
where
\begin{equation}
\Delta\theta=\theta(\mathbf r_0+\Delta\mathbf r,t_0+\Delta t)-\theta(\mathbf r_0,t_0).
\end{equation}
The KPZ scaling hypothesis gives
\begin{equation}
G_2(\Delta r,\Delta t)=\Delta t^{2\beta}\mathcal{F}\left(\frac{\Delta r}{\Delta t^{1/z}}\right),
\end{equation}
with $\mathcal{F}(x)\sim x^{2\chi}$ in the KPZ regime. As is evident in Fig.~\ref{fig:analytic}d, the rescaled correlators for different $\Delta t$ collapse onto a common, universal curve over the accessible scaling window. This data collapse provides a direct benchmark for the corresponding open-GPE analysis below.

\begin{figure}[t]
\centering
\includegraphics[width=\linewidth]{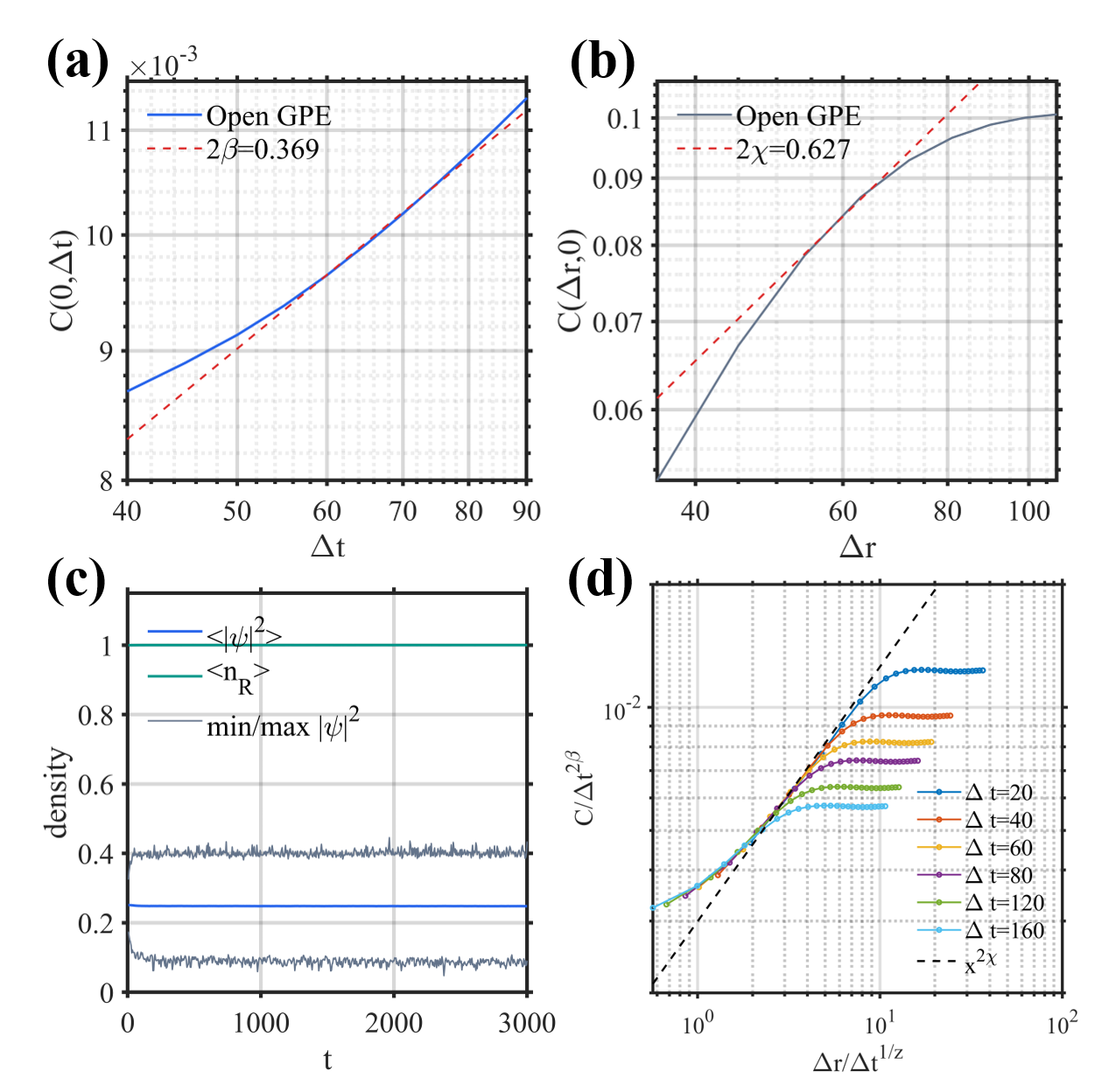}
\caption{Disordered-pump stochastic open-GPE calculation. The run uses $L=128$, $dx=9$, $dt=0.25$, $t_{\rm max}=3000$, $P_0=1.05P_{\rm th}$, weak annealed gain disorder $\sigma_P=0.0018$, $\xi_P=18$, and $\Delta t_P=1$. The calibrated target phase coefficients are $\nu=0.005$, $|\lambda|=1.25$, and $D_{\rm eff}=0.1$. (a) The compact temporal coherence gives slope $0.367$ over $50\le\Delta t\le80$, close to the $2\beta=0.369$ benchmark. (b) Compact spatial coherence. (c) The density and reservoir remain stable during the run. We choose $t_0=1~{\rm ps},\  \ell_0=0.962~\mu{\rm m},\ n_0=100~\mu{\rm m}^{-3}$ to translate the dimensionless open-GPE parameters into SI quantities.  (d) Joint correlation function for the same parameter set. We compute $C(\Delta r,\Delta t)=-\ln|g^{(1)}(\Delta r,\Delta t)|$ from complex open-GPE snapshots and plot $C/\Delta t^{2\beta}$ against $\Delta r/\Delta t^{1/z}$. The curves for $\Delta t=20,40,60,80,120,160$ collapse in the KPZ variables, supporting the dynamic scaling form of Eq.~\eqref{eq:g1_scaling} over the accessible window. Other corresponding parameters in SI are $m=1.0\times10^{-4}m_e$, $\gamma_c=1.0~{\rm ps}^{-1}$, $\gamma_R=5.0~{\rm ps}^{-1}$, $R=1.0\times10^{-2}~\mu{\rm m}^3{\rm ps}^{-1}$, $g=6.6\times10^{-3}~{\rm meV}\,\mu{\rm m}^{3}$, $g_R=3.45\times10^{-2}~{\rm meV}\,\mu{\rm m}^{3}$, $P=525~\mu{\rm m}^{-3}{\rm ps}^{-1}=1.05P_{\rm th}$, $D_R=9.3\times10^{-2}~\mu{\rm m}^2{\rm ps}^{-1}$, $\nu=4.6\times10^{-3}~\mu{\rm m}^2{\rm ps}^{-1}$, $|\lambda|=1.16~\mu{\rm m}^2{\rm ps}^{-1}$, $D_{\rm eff}=8.9\times10^{-2}~\mu{\rm m}^{3}{\rm ps}^{-1}$, $\sigma_P=1.8\times10^{-3}$, $\xi_P=17~\mu{\rm m}$, $\Delta t_P=1~{\rm ps}$, $dx=8.7~\mu{\rm m}$, $dt=0.25~{\rm ps}$, $t_{\rm max}=3~{\rm ns}$.}
\label{fig:numeric}
\end{figure}

Having established the temporal, spatial, and full dynamical scaling of the $3+1$-dimensional KPZ control problem, we next examine whether the same correlation structure emerges from the full stochastic
open-dissipative polariton model in Eq.~\eqref{eq:gpe}. The simulations are performed slightly above the condensation threshold, where the condensate density is large enough to suppress amplitude defects but the effective KPZ coupling remains strong. Uniform pumping is an important negative control for the microscopic model. In a large, stable three-dimensional condensate, it can produce long-range off-diagonal order, so the compact first-order coherence may remain nearly flat even though the KPZ equation would already show roughening. To mimic stochastic growth more directly in the open-GPE model, we therefore replace the strictly uniform gain by a weakly disordered extended pump,
\begin{equation}
 P({\bf r},t)=P_0\exp[\sigma_P u({\bf r},t)-\sigma_P^2/2],
 \label{eq:disordered_pump}
\end{equation}
where $u$ is a zero-mean Gaussian field filtered over a spatial scale $\xi_P$ and refreshed on a time scale $\Delta t_P$. The normalization keeps the spatial average close to $P_0$.

Figure~\ref{fig:numeric} shows the long-time check of open-GPE. The compact temporal coherence gives
\begin{equation}
 C(0,\Delta t)\propto \Delta t^{0.367}
 \quad (50\le\Delta t\le80),
\end{equation}
which is close to the $3+1$ KPZ value $2\beta=0.369$.
The compact spatial coherence gives
\begin{equation}
 C(\Delta r,0)\propto \Delta r^{0.627}
 \quad (45\le \Delta r\le81),
\end{equation}
which is identical to $2\chi=0.627$.

The condensate density stays finite throughout the run, with final mean density $\langle|\psi|^2\rangle=0.248$ and global minimum $|\psi|^2=0.047$. We therefore interpret this result as a useful operating window in which the full microscopic open-GPE model exhibits three-dimensional KPZ scaling. The strongest check is obtained from the full spatial-temporal correlation function. For the same open-GPE parameters, we calculated the evolution of polariton condensate and evaluated the normalized correlation function in Eq.~\eqref{eq:g1_def}. Figure~\ref{fig:numeric}(d) shows the corresponding
\begin{equation}
 C(\Delta r,\Delta t)=-\ln|g^{(1)}(\Delta r,\Delta t)|
\end{equation}
plotted in the KPZ variables $C/\Delta t^{2\beta}$ and $\Delta r/\Delta t^{1/z}$. The collapse of the curves for different $\Delta t$ supports the dynamic scaling form in Eq.~\eqref{eq:g1_scaling}. Because the collapse is obtained in a finite time and size window, the asymptotic convergence is a strong evidence for an accessible KPZ scaling regime.

\begin{figure}[t]
\centering
\includegraphics[width=\linewidth]{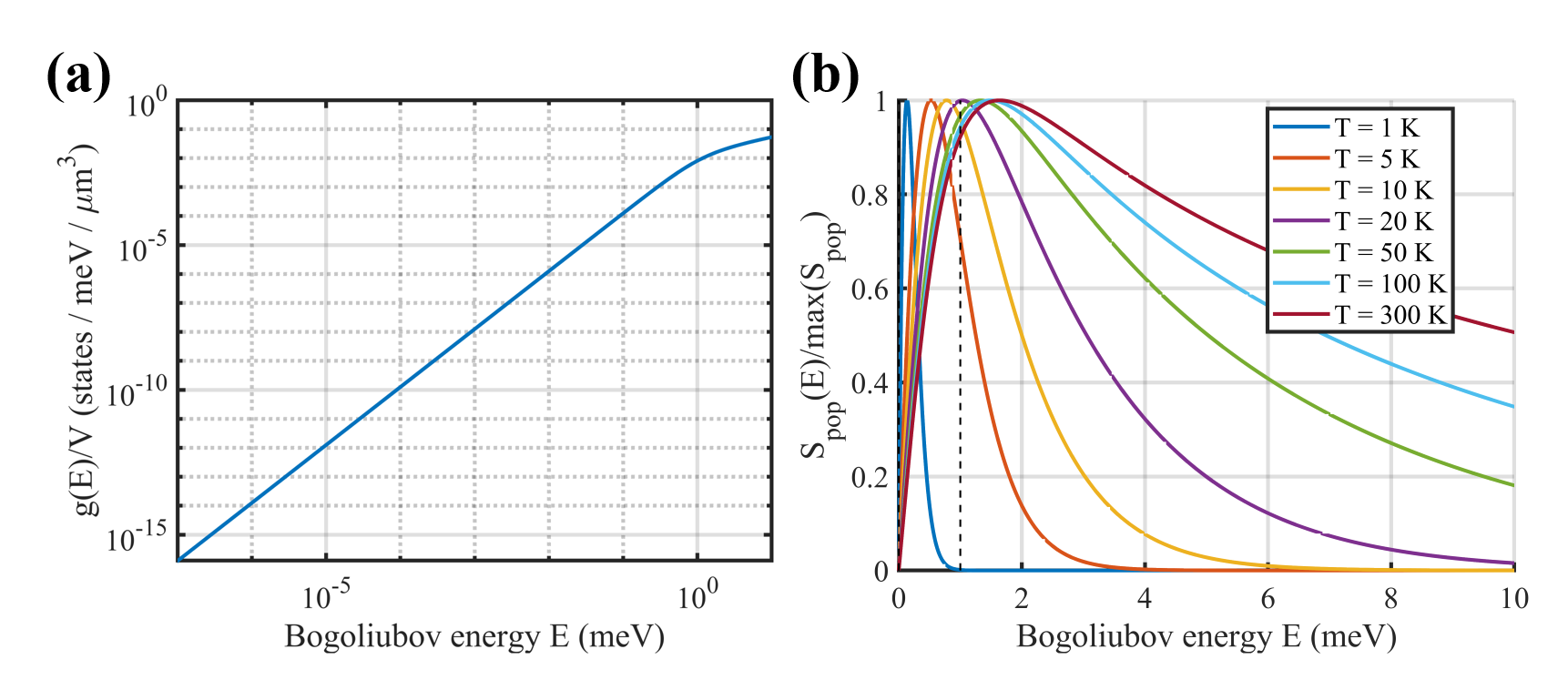}
\caption{
(a) Density of states $g_{\rm 3D}(E)$ of 3D Bogoliubov dispersion. (b) Normalized energy-resolved population spectrum of three-dimensional
Bogoliubov excitations,
$S_{\rm pop}(E)=g_{\rm 3D}(E)f_{\rm B}(E,T_{\rm eff})$, for several effective temperatures $T=1,5,10,20,50,100,300\ \rm K$.
 Chemical potential $\mu$ is set as $1$ meV.}
\label{fig:efftemp}
\end{figure}

The stochastic noise entering Eqs.~\eqref{eq:gpe} and ~\eqref{eq:reservoir} characterizes a local and Markovian driving source, whereas the response of the condensate is resolved through its collective excitation modes. In particular, the long-wavelength fluctuations of a three-dimensional condensate are carried by the Bogoliubov Goldstone mode,
\begin{equation}
E_{\rm B}(\mathbf{k})
=
\sqrt{E_0(\mathbf{k})
\left[E_0(\mathbf{k})+2\mu\right]},
\end{equation}
where
\begin{equation}
E_0(\mathbf{k})
=
\sum_{i=x,y,z}\alpha_i k_i^2,
\qquad
\alpha_i=\frac{\hbar^2}{2m_i},
\end{equation}
and $\mu=g\rho_0$. If the occupation of these modes is phenomenologically parametrized by an effective Bose distribution, the energy-resolved population spectrum can be written as
\begin{equation}
S_{\rm pop}(E)
=
g_{\rm 3D}(E)
f_{\rm B}(E,T_{\rm eff}),
\end{equation}
where $g_{\rm 3D}(E)$ is the Bogoliubov density of states and $f_{\rm B}(E,T_{\rm eff})= [\exp(E/k_{\rm B}T_{\rm eff})-1]^{-1}$. In the Goldstone regime, $E\ll\mu$, the linear Bogoliubov dispersion gives $g_{\rm 3D}(E)\propto E^2$. Consequently, in the classical low-energy limit $E\ll k_{\rm B}T_{\rm eff}$, $S_{\rm pop}(E)\propto E$.
Thus, unlike in two dimensions, a three-dimensional Goldstone mode does not produce a low-energy white spectral plateau. Figure~\ref{fig:efftemp} illustrates the 3D Bogoliubov density of states and this crossover for several phenomenological values of $T_{\rm eff}$.

The polariton-crystal platform addresses the main obstacles that have limited experimental access to higher-dimensional KPZ scaling. First, the condensate forms in a Bloch band of a three-dimensional photonic crystal instead of in the plane of a Fabry--P\'erot cavity. This avoids reducing the long-wavelength phase dynamics to a two-dimensional problem. Second, the band curvature near the condensation valley is an engineering parameter. By choosing the exciton detuning, the light-matter coupling, and the symmetry-selected photonic branch, one can tune the effective mass tensor and hence the coefficients $\nu_i$ and $\lambda_i$ in Eq.~\eqref{eq:anisotropic_kpz}. Third, the system is  driven and dissipative. Pump and loss are not parasitic perturbations; they are the physical ingredients that generate the KPZ phase dynamics.

The proposed measurement follows the standard logic of polariton coherence experiments, generalized to a three-dimensional photonic crystal. Momentum-resolved photoluminescence identifies the condensation valley and verifies that the emission comes from the three-dimensional lower-polariton branch. Interferometric measurements of the emitted field give the temporal coherence $\gone(0,\Delta t)$ and projected spatial coherences along selected crystal directions. Measurements on different facets, or tomographic reconstruction of the leakage from symmetry-related directions, can test the isotropy of the scaling and reconstruct the full three-dimensional correlation function. The most direct evidence for $3+1$ KPZ scaling is the simultaneous consistency of temporal and spatial exponents with Eq.~\eqref{eq:3d_exponents}, together with the scaling collapse in Eq.~\eqref{eq:g1_scaling}.

There are also clear experimental caveats. A finite crystal thickness produces eventual saturation. Disorder can localize low-energy modes if it exceeds the phase stiffness. Strong valley anisotropy may generate an anisotropic KPZ crossover before the isotropic fixed point is reached. Excessive pumping suppresses phase fluctuations and can hide the KPZ regime, whereas pumping too close to threshold can generate amplitude defects. These constraints define the optimal operating window: a large three-dimensional crystal, stable condensation, pump moderately above threshold, and sufficient noise and nonlinearity to place the phase dynamics on the strong-coupling side of the three-dimensional KPZ transition.


\paragraph*{Conclusion.---}
We have shown that a three-dimensional polariton crystal can realize the phase dynamics required for $3+1$ KPZ scaling. The essential ingredients are a genuine three-dimensional lower-polariton condensate, driven-dissipative phase dynamics, a tunable effective mass tensor, and an experimentally accessible first-order coherence function. Eliminating the massive density and reservoir modes gives an effective KPZ equation for the condensate phase. Numerical simulations of both the equation and the full driven-dissipative polariton model exhibit an intermediate-asymptotic regime consistent with the $3+1$ KPZ exponents. The result provides a concrete route to nonequilibrium universality beyond two dimensions and identifies three-dimensional polariton crystals as a platform where photonic band engineering and stochastic many-body scaling can be studied in the same experiment.

\paragraph*{Acknowledgments.---}
AVK acknowledges support from Saint Petersburg State University (Research Grant No. 125022803069-4) and from the Innovation Program for Quantum Science and Technology (No. 2021ZD0302704). AA acknowledges support from the Russian Science Foundation grant No. 25-11-00114.

\bibliography{apssamp}

\end{document}